%
%
%

\ifx\mnmacrosloaded\undefined \input mn\fi

%

\newif\ifAMStwofonts

\ifCUPmtplainloaded \else
  \NewTextAlphabet{textbfit} {cmbxti10} {}
  \NewTextAlphabet{textbfss} {cmssbx10} {}
  \NewMathAlphabet{mathbfit} {cmbxti10} {} 
  \NewMathAlphabet{mathbfss} {cmssbx10} {} 
  \ifAMStwofonts
    \NewSymbolFont{upmath} {eurm10}
    \NewSymbolFont{AMSa} {msam10}
    \NewMathSymbol{\upi}     {0}{upmath}{19}
    \NewMathSymbol{\umu}     {0}{upmath}{16}
    \NewMathSymbol{\upartial}{0}{upmath}{40}
    \NewMathSymbol{\leqslant}{3}{AMSa}{36}
    \NewMathSymbol{\geqslant}{3}{AMSa}{3E}

  \else
    \def\umu{\mu}
    \def\upi{\pi}
    \def\upartial{\partial}
  \fi
\fi



\loadboldmathnames



\letters          
\pagerange{L00--L00}    
\pubyear{1999}
\volume{000}

\begintopmatter  

\title{Are there any Type~2 QSOs? The case of AXJ0341.4--4453}

\author{J. P. Halpern$^1$, T. J. Turner$^{2,3}$ and I. M. George$^{2,4}$}
\vskip 1pt
\affiliation{$^1$Columbia Astrophysics Laboratory, Columbia University,
550 West 120th Street, New York, NY 10027, USA}
\vskip 3pt
\affiliation{$^2$Laboratory for High Energy Astrophysics, Code 660,
NASA Goddard Space Flight Center, Greenbelt, MD 20771, USA}
\vskip 3pt
\affiliation{$^3$University of Maryland, Baltimore County}
\vskip 3pt
\affiliation{$^4$Universities Space Research Association}

\shortauthor{J. P. Halpern et al.}
\shorttitle{Are there any Type~2 QSOs?}


\acceptedline{Accepted . Received ; in original form}

\abstract {The X-ray source AXJ0341.4--4453 was described by Boyle et al.
as a Type~2 AGN at $z = 0.672$ based on the absence of broad emission lines
in the observed wavelength range 4000--7000 \AA.  We obtained a new spectrum 
of AXJ0341.4--4453 extending to 9600~\AA\ which reveals
broad Balmer lines and other
characteristics of Seyfert~1 galaxies.  The FWHM of broad H$\beta$
is at least 1600 km~s$^{-1}$, while [O~{\sc iii}]~$\lambda$5007 has FWHM = 730
km~s$^{-1}$.  The flux ratio [O~{\sc iii}]~$\lambda$5007/H$\beta$ = 1.
Thus, AXJ0341.4--4453 is by definition a narrow-line Seyfert~1 galaxy,
or perhaps a moderately reddened Seyfert~1 galaxy, 
but it is not a Type~2 QSO.
Although examples of the latter have long been sought, particularly in
connection with the problem of the X-ray background, there is still
virtually no evidence for the existence of {\it any} 
Type~2 QSO among X-ray selected samples.}


\maketitle  

\section{Introduction}

The X-ray source AXJ0341.4--4453 was discovered in a deep exposure by
the {\sl ASCA\/} satellite, and was noted to have an exceptionally hard
X-ray spectrum (Boyle et al. 1998). It was only weakly detected in a
deep {\sl ROSAT\/} survey of the same field (Georgantopoulos et al. 1996).
An optical identification was
made with an emission-line object at $z = 0.672$
that has high-ionization forbidden lines,
most notably [Ne~{\sc v}]~$\lambda$3426 (Boyle et al. 1998).
The absence of broad Mg~{\sc ii}~$\lambda$2798 and broad Balmer lines
led those authors to classify AXJ0341.4--4453 as a Type~2 (obscured)
AGN with a 2--10 keV X-ray luminosity of $1.8 \times 10^{44}$ erg~s$^{-1}$.

It has long been postulated that highly absorbed AGNs are
the source of most of the hard X-ray background.  The search for
Type~2 QSOs among X-ray survey identifications is therefore
a natural test of this hypothesis,
but there is little, if any,
evidence for such high-luminosity analogues of Seyfert~2
galaxies (see Halpern, Eracleous \& Forster 1998; Halpern \& Moran 1998; 
and references therein).
Because they are rare (possibly nonexistent) and easily 
mimicked by other types of AGNs, considerable care must
be taken in the search for true Type~2 QSOs.
In particular, the
classification of AXJ0341.4--4453 as Type~2 AGN was made on
the basis of an optical spectrum extending to a maximum
wavelength of 4200 \AA\ in
the rest frame, thus excluding the lower order Balmer lines
from examination.  Since these lines are sometimes the only
broad lines that are detected in X-ray selected AGNs, we sought
to remedy this deficiency by obtaining a spectrum of
AXJ0341.4--4453 extending at least to the H$\beta$ and
[O~{\sc iii}]~$\lambda$5007 emission lines.

\section{Observations}

Optical spectra of AXJ0341.4--4453 were obtained on UT 1998 October~19
using the RC Spectrograph and Loral 3K CCD on the CTIO 4m telescope.
The wavelength range 3600--9600 \AA\ was covered at a resolution of
6.3~\AA.  Wavelength calibration using a helium-argon spectrum
achieved an rms accuracy of 0.3~\AA.
Both object and standard star LTT~1788 (Baldwin \& Stone 1983,1984)
were observed near zenith under photometric conditions
through a slit of width $1.\!\!^{\prime\prime}5$.
A total of 3000~s of exposure was obtained on AXJ0341.4--4453.
The spectral images were reduced using standard methods and extracted
using an optimal extraction technique (Horne 1986).
The reduced spectrum is shown in Figure~1.  Although a WG360 blocking 
filter was used, there is still the possibility of second-order overlap
at wavelengths longer than 7200~\AA.  However, both object and standard star
are relatively red, and we see no specific evidence for
second-order contamination in either spectrum. The standard star was also
used as a template for the removal of atmospheric absorption bands in the
red.   Instrumental artefacts are apparent near wavelengths 6860~\AA\ and
8870~\AA.  These are due to charge traps in the CCD that are approximately
10 columns (20 \AA) wide.
\input psfig.sty
\beginfigure*{1}
\psfig{figure=/apian3/jules/jane/fig_1.ps,angle=-90.,height=6.65cm}
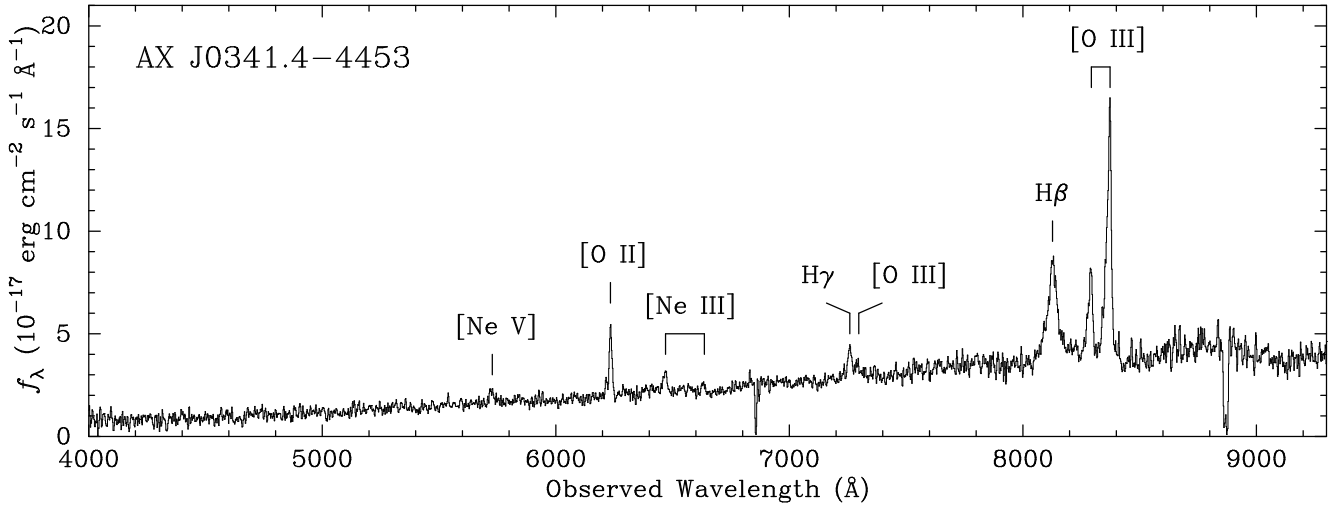
\caption{{\bf Figure 1.} Spectrum of AXJ0341.4--4453 from the CTIO
4m telescope on UT 1998 October 19.}
\endfigure
\beginfigure{2}
\psfig{figure=/apian3/jules/jane/fig_2.ps,angle=-90.,height=6.0cm}
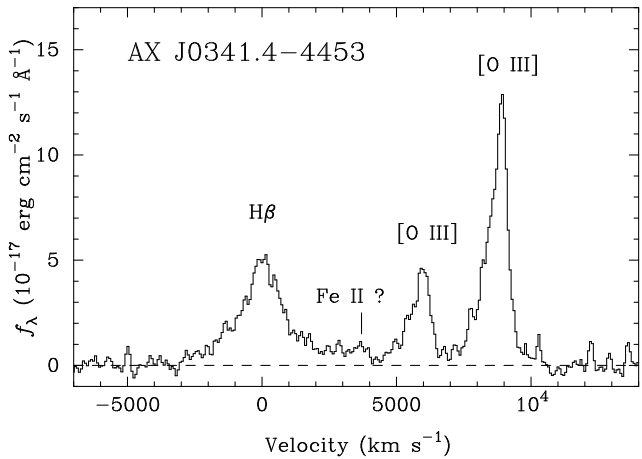
\caption{{\bf Figure 2.} Continuum subtracted spectrum of AXJ0341.4--4453
in velocity units.  A feature possibly due to Fe~{\sc ii}~$\lambda$4923 is
marked.}
\endfigure
\beginfigure{3}
\psfig{figure=/apian3/jules/jane/fig_3.ps,angle=-90.,height=6.0cm}
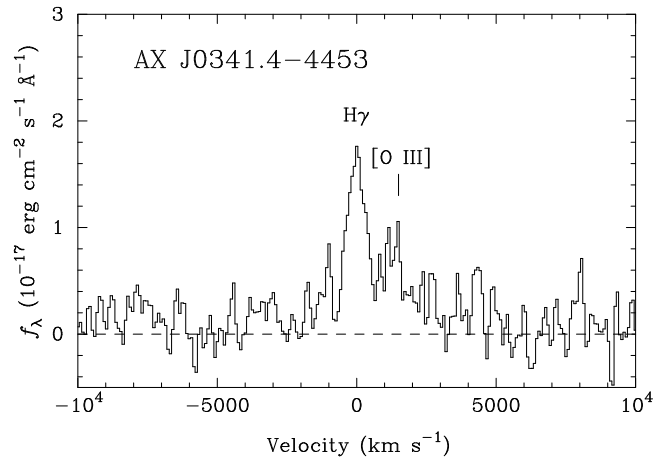
\caption{{\bf Figure 3.} Same as Figure 2, for the H$\gamma$ region.}
\endfigure

Our spectrum resembles that of Boyle et al. (1998) shortward of
7000 \AA, but we also cover the redshifted
emission lines of H$\gamma$, H$\beta$,
and [O~{\sc iii}] in the near infrared.  The Balmer lines contain components
that are broader than the forbidden lines, immediately requiring
a Seyfert~1 classification.   Details of the profiles can be seen
in Figures~2 and~3, which show the regions around the
Balmer lines in velocity units, after continuum
subtraction.
The [O~{\sc iii}] line has FWHM = 730 km~s$^{-1}$, while the FWHM of the total
H$\beta$ profile is 1620 km~s$^{-1}$, more than twice that of [O~{\sc iii}].
The signal-to-noise ratio is not sufficient to decompose
H$\beta$ into narrow and broad components,
but it is clear that the H$\beta$ line extends to
$\pm 3000$ km~s$^{-1}$, whereas the [O~{\sc iii}] lines have no such component.
In Figure 1 there is a hint of the presence of the usual
complexes of permitted Fe~{\sc ii} multiplets on
either side of H$\beta$ and [O~{\sc iii}].  Although these are too noisy to measure,
it is possible, for example, that Fe~{\sc ii}~$\lambda$4923 contributes to an
apparent emission bump at 3000--4000 km~s$^{-1}$ in the rest
frame of H$\beta$ as indicated in Figure~2.
Note that H$\delta$ in our spectrum falls on the charge trap
at 6860 \AA, and is therefore not detected.   H$\delta$ was the only Balmer
line visible in the spectrum of Boyle et al. (1998), and its blue side
in fact looks somewhat broad there.  However, the red side of the H$\delta$
profile is absorbed by the atmospheric B band at 6867 \AA, which was evidently
not corrected in their spectrum, and is responsible for their discrepant
redshift measurement from this line.

\begintable{1}
\caption{{\bf Table 1.} Emission-line measurements of AXJ0341.4--4453.}
\halign{%
\rm#\hfil & \qquad\rm\hfil# & \quad\rm\hfil# & \qquad\rm\hfil#\hfil \cr
\noalign{\vskip 5pt \hrule \vskip 5pt}
Line & \omit\hfil Flux\hfil & FWHM & $z$ \cr
& (erg cm$^{-2}$ s$^{-1}$) & (km s$^{-1}$) & \cr
\noalign{\vskip 5pt \hrule \vskip 5pt}
Mg~{\sc ii}~$\lambda$2798              & $< 2 \times 10^{-16}$ & ...  & ...    \cr
[Ne~{\sc v}]~$\lambda$3426             & $1.3 \times 10^{-16}$ & 1060 & 0.6709    \cr
[O~{\sc ii}]~$\lambda$3727             & $4.7 \times 10^{-16}$ & 630  & 0.6725 \cr
[Ne~{\sc iii}]~$\lambda$3869           & $1.9 \times 10^{-16}$ & 860  & 0.6718   \cr
[Ne~{\sc iii}]~$\lambda$3968, H$\epsilon$ & $6.8 \times 10^{-17}$ & ...  & ...    \cr
H$\gamma$ (narrow)                     & $2.0 \times 10^{-16}$ & 740  & 0.6724    \cr
H$\gamma$ (broad)                      & $4.6 \times 10^{-16}$ & ...  & ...    \cr
[O~{\sc iii}]~$\lambda$4363            & $1.0 \times 10^{-16}$ & ...  & ...    \cr
H$\beta$ (total)                       & $3.1 \times 10^{-15}$ & 1620 & 0.6719 \cr
[O~{\sc iii}]~$\lambda$4959            & $1.2 \times 10^{-15}$ & 720  & 0.6718 \cr
[O~{\sc iii}]~$\lambda$5007            & $3.3 \times 10^{-15}$ & 730  & 0.6723  \cr
\noalign{\vskip 5pt\hrule}
}
\endtable

Details of our emission-line measurements are given in Table~1.
We measure a redshift of $z = 0.6723 \pm 0.0002$ from the stronger
narrow emission lines, in agreement with Boyle et al. (1998).
We do not list dereddened fluxes, as Galactic extinction is
negligible in this direction.
Although it is not possible to decompose the H$\beta$ line uniquely
into broad and narrow components, the H$\gamma$ line has a narrow
peak which is more readily measured (see Figure 3), thus providing evidence
that the narrow-line region is less reddened than the region emitting
the broad lines.  We give only the total H$\beta$ flux in Table~1,
but we list the broad and narrow components of H$\gamma$ separately.
If we {\it assume} that the narrow-line ratio
H$\gamma$/H$\beta$ has the recombination value 0.47, then the broad
H$\gamma$/H$\beta$ ratio is 0.17.  Such a decomposition of the
H$\beta$ flux is plausible, since it would require
[O~{\sc iii}]~$\lambda$5007/H$\beta = 8$
for the narrow-line region, a value which is typical.
According to the standard Galactic reddening law,
H$\gamma$/H$\beta = 0.17$ would correspond to 7 magnitudes of
visual absorption intrinsic to the broad-line region.
However, we caution that the decomposition of the H$\gamma$ line is
very uncertain, and sensitive to the placement of the continuum.
Our estimate of $A_V$ is probably an upper limit, as the narrow-line region
could be somewhat reddened as well, and the broad H$\gamma$/H$\beta$ ratio
could also be affected by collisional excitation and radiative transfer effects,
and contamination of H$\beta$ by Fe~{\sc ii}. 
Extinction could also be in part
responsible for the absence (noted by Boyle et al. 1998)
of Mg~{\sc ii}~$\lambda$2798,
which is not seen at the expected wavelength of 4680 \AA.
For a typical Mg~{\sc ii}/H$\beta$ ratio of 1, $A_V = 3.5$
would be required to reduce Mg~{\sc ii} to below our upper limit.

\section{Conclusions and Speculations}

The optical spectrum of AXJ0341.4--4453 is most naturally interpreted as
that of a
narrow-line Seyfert~1 galaxy (NLS1). It fits the standard definition
of this class (Osterbrock \& Pogge 1985; Goodrich 1989), namely
[O~{\sc iii}]~$\lambda$5007/H$\beta < 3$ {\it and}
FWHM H$\beta < 2000$ km~s$^{-1}$.
There is also possible evidence for permitted Fe~{\sc ii} emission lines,
which are common in NLS1s.  The only uncertainty in classification
involves the extent of the broad component of H$\beta$,
which could be larger than we have detected.  If the narrow and broad
components could be separated in spectra of higher signal-to-noise
ratio, then the FWHM of the broad component alone could be larger,
which might require revision to an ordinary Seyfert 1 classification.
In NLS1s, however, these components are normally not separable. 
In any case, AXJ0341.4--4453 certainly does not have a Seyfert 2 spectrum,
and is therefore not a candidate for the high-luminosity Seyfert~2 analogue,
the Type~2 QSO.  It is easy to mistake a NLS1 for a Type~2 QSO.
Another such case was
IRAS 20181--2244 (Halpern \& Moran 1998), a NLS1 that also suffers
from moderate absorption, and lacks a Mg~II emission line.

Moderate obscuration of the broad-line and continuum emitting regions can
probably account for the optical and X-ray properties of AXJ0341.4--4453.
A column density of $\sim 10^{22}$~cm$^{-2}$, which severely attenuates
soft X-rays, could be responsible for the apparent hard X-ray spectrum,
as well as for several magnitudes of visual extinction if the dust and
gas properties are similar to Galactic.  Although Boyle et al. (1998)
dismissed the NLS1 possibility on the basis of a hard X-ray spectrum,
it is now evident from their deep X-ray survey, as well as others,
that the types of AGNs found at faint flux levels in hard
X-ray surveys with {\sl ASCA\/} and {\sl BeppoSAX\/}
(Akiyama et al. 1998; Fiore et al. 1999;
Giommi, Fiore \& Perri 1998) are basically the same as had been found 
previously in soft X-rays with {\sl ROSAT}.
Therefore, it should not be surprising that moderately
obscured but otherwise representative AGNs are found
when X-ray sources are singled out on the basis of their flat X-ray spectra.
We also caution that, since even highly luminous NLS1s can
vary by more than order of magnitude in X-ray flux
(e.g., Forster \& Halpern 1996), 
it is not safe to draw conclusions about X-ray spectral shape by comparing
noncontemporaneous {\sl ROSAT\/} and {\sl ASCA\/}
observations of these objects.

As to the question of whether or not such moderately reddened objects
{\it are} Type~2 AGNs, we remark that
reddening alone is not sufficient to make a Type~2 AGN.  It depends
on where the obscuring material is located and how thick it is.
In the currently popular unified model of Seyfert classification,
the obscuring material must be in a position to obscure the
broad-line emitting region from our view, but not the narrow-line region,
in order to produce a Type~2 AGN.  A visual absorption of 3.5--7 magnitudes,
as estimated above, is vanishingly small compared to the 
amount of extinction in the so-called Compton-thick Seyfert~2 galaxies,
for which equivalent visual absorption of 1000 magnitudes or more is inferred 
from their X-ray measured column densities.  An absorption of 3.5--7 magnitudes 
could easily be due to the interstellar medium
in the disk of the host galaxy, especially if 
viewed at high inclination, or to a single molecular cloud.
Indeed, a certain fraction of Seyferts {\it must}
suffer just such galaxian obscuration. Therefore, isolated 
cases like AXJ0341.4--4453 cannot be interpreted in the unified 
model as Type~2, especially since Compton-thick absorption is now
found to be ubiquitous among Seyfert~2 samples that are selected by their
[O~{\sc iii}] flux alone (Maiolino et al. 1998). AXJ0341.4--4453 shows
no evidence of being Compton thick.  In particular, its
ratio $L_{\rm X}/L([{\rm O}~{\sc iii}])$ of 33 is much larger than those of the 
Compton-thick Seyferts in Maiolino et al. (1998), for which this ratio
is less than 1.

It would also be unpalatable to regard a NLS1
as an ``intermediate'' type AGN. The recently acquired wealth of data on
NLS1s eludes explanation within the Seyfert unification
paradigm.  In particular, the fact that the permitted emission lines
are narrow in NLS1s is not understood in terms of any
orientation-dependent unification scheme.  Rather, it is
likely to be due to some intrinsic physical property.
Similarly, the rapid and large amplitude X-ray variability,
and soft X-ray spectra which characterise NLS1s as a class,
are antithetical to obscuration.

The absence of Type~2 QSOs, the high-luminosity analogues of Seyfert~2
galaxies, remains a significant fact to be explained whether in the
context of unified models or not.  Broad optical emission
lines are detectable in most Seyfert galaxies with X-ray luminosity
$> 10^{42}$ erg~s$^{-1}$ (Halpern, Helfand \& Moran 1995;
Moran, Halpern \& Helfand 1996),
and, as far as we know, in {\it every} non-blazar
AGN whose X-ray luminosity exceeds $3 \times 10^{44}$ erg~s$^{-1}$
(Halpern et al. 1998). In the context of the unified scheme, the
the absence of Type~2 QSOs among X-ray selected samples is natural
if either (1) all such objects are perfectly Compton thick or (2)
all sufficiently luminous QSO nuclei are able to remove any obscuring
material from their vicinity, affording their broad-line regions
$4\pi$ steradians of visibility.  But we do not have much confidence
that either of these idealizations will hold strictly true.  Therefore,
it is possible that Type~2 QSOs may yet be
found at the lower X-ray flux thresholds of {\sl AXAF\/} and {\sl XMM\/}.
However, it is not clear that such a new population is needed to account
for the hard X-ray background.  Indeed, the ideal Compton-thick AGN can
make little or no contribution to the X-ray background unless 
such a source is at high redshift.  Whether or not Type~2 QSOs exist,
it might turn out that moderately obscured
AGNs of the types that are well known will prove sufficient to
comprise the X-ray background (Fiore et al. 1999; Giommi et al. 1998).

\section*{Acknowledgments}

The authors were visiting astronomers
at Cerro-Tololo Inter-American Observatory,
National Optical Astronomy Observatories, which is operated by AURA, Inc.,
under a cooperative agreement with the US National Science Foundation.
We thank Mauricio Navarrete for user support at CTIO.


\bye